\documentclass[aps,prb,twocolumn,showpacs,floatfix]{revtex4}
\usepackage{graphicx}
\usepackage{bm}
\usepackage{color}
\newcommand{\be}{\begin{equation}}
\newcommand{\ee}{\end{equation}}
\newcommand{\bk}{{{\bf{k}}}}

\newcommand{\br}{{{\bf{r}}}}

\newcommand{\bea}{\begin{eqnarray}}
\newcommand{\eea}{\end{eqnarray}}
\newcommand{\ra}{\rangle}

\newcommand{\dg}{{\dagger}}
\newcommand{\pdg}{{\phantom\dagger}}

\begin{document}

\title{High-temperature large-gap quantum anomalous Hall insulator in ultrathin double perovskite films}

\author{Santu Baidya$^{1,\ast}$, Umesh V. Waghmare$^{2}$, Arun Paramekanti$^{3,4}$, Tanusri Saha-Dasgupta$^{5,\dagger}$}
\affiliation{$^{1}$ Department of Physics and Center for Nanointegration Duisburg-Essen (CENIDE), University of Duisburg-Essen,
Lotharstrasse 1, 47057 Duisburg, Germany. \\
$^{2}$ Jawaharlal Nehru Centre for Advanced Scientific Research,  Jakkur, Bangalore 560064, India. \\
$^{3}$ Department of Physics, University of Toronto, Toronto, Ontario, Canada M5S 1A7. \\
$^{4}$ Canadian Institute for Advanced Research, Toronto, Ontario, M5G 1Z8, Canada. \\
$^{5}$ Department of Condensed Matter Physics and Materials Science, S. N. Bose National Centre for Basic Sciences, Kolkata 700 098, 
India.}

\date{\today}

\begin{abstract}
Motivated by the goal of realizing topological phases in thin films and heterostructures of correlated oxides, 
we propose here a quantum anomalous Hall insulator (QAHI) in ultrathin 
films of double perovskites based on mixed 3$d$-5$d$ or 3$d$-4$d$ transition metal ions, grown along [111] direction. Considering the specific case of ultrathin Ba$_2$FeReO$_6$, we present a theoretical analysis of an 
effective Hamiltonian derived from first-principles. We establish that a strong spin-orbit coupling
at Re site, $t_{2g}$ symmetry of the low-energy $d$-bands, polarity of its [111] orientation of perovskite
structure, and mixed 3$d$-5$d$ chemistry results in room temperature magnetism with a robust QAHI
state of Chern number $C$ = 1 and a large band-gap.
We uncover and highlight a non-relativistic {\it orbital Rashba}-type 
effect in addition to the spin-orbit coupling, that governs this QAHI state. 
Our prediction of a large topological band-gap
of $\sim 100$ meV in electronic structure, and a magnetic transition temperature
$T_c \sim 300$K estimated by Monte Carlo simulations, is expected to stimulate experimental efforts
at synthesis of such films and enable possible practical applications of its dissipationless edge currents.
\end{abstract}

\maketitle

\section{Introduction}

Quantum anomalous Hall insulators are two-dimensional (2D) band insulators exhibiting a quantized Hall conductivity
\cite{Haldane88,CXLiu15} in the presence of magnetic order that spontaneously breaks time-reversal symmetry. The requisite intrinsic nonzero Berry curvature \cite{TKNN} for such anomalous Hall effect 
could appear either due to the adiabatic motion of electrons in magnetically ordered backgrounds with nontrivial pinned 
spin-textures \cite{Martin08,Kumar12,Motome12,Tokura12,MacDonald14} or from spin-orbit coupling (SOC) even in simple
collinear ferromagnetic (FM) or antiferromagnetic states.\cite{Onoda03,CXLiu08,Niu14}

Following a theoretical proposal \cite{ZhongFang10}, the first observations of QAHIs have 
been made in topological insulator (TI) thin films doped with magnetic Cr or V atoms, 
with a FM $T_c$ of $\sim$ 15$-$20 K, and a quantized $\sigma_{xy} \!=\! e^2/h$ 
observed for $T \! \lesssim \! 0.5$K.\cite{Xue13,Moodera15}
There have been other proposals for QAHIs, yet to be realized, via depositing magnetic 
atoms on graphene \cite{Niu10,Mokrousov12} or by putting heavy atoms with large 
SOC on magnetic substrates.\cite{Vanderbilt13} These proposals share a common feature of 
topologically nontrivial bands derived from $s$ or $p$ orbitals. Given the potential 
applications of chiral edge currents associated with QAHI in dissipationless electronic 
circuits,\cite{CXLiu15} its desirable for practical applications
that  one would like to i) boost the temperature scale for QAH effect, ii) engineer a 
large topological band-gap, and iii) avoid possible dopant or adatom-induced inhomogeneity 
by considering stoichiometric composition. 

Considering the recent progress in growing oxide heterostructures along different crystallographic 
directions,\cite{Freeland10,Gibert12,Catalano15,Hirai15,Middey16} and the drive towards integrated oxide 
electronics,\cite{Hwang12,Schlom15} ultrathin films of transition metal oxides (TMOs) appear to be 
an attractive possibility in the above context. Compared to $s$ or $p$ orbitals based systems, TMOs with
$d$ orbital dominated band structure has the possibility of high temperature realization. 
Seminal theoretical studies \cite{Xiao11,Ran11} on perovskite bilayers grown 
along the [111] crystallographic direction predicted that the combination of the buckled honeycomb 
structure and SOC can lead to 2D time-reversal invariant topological insulator in $t_{2g}$ bands 
in SrIrO$_3$ and LaOsO$_3$ bilayers, as well as interaction-induced stabilization of QAHIs 
in flat $e_g$ bands of LaAuO$_3$ bilayers. Following this,
ultrathin films hosting nickelate $e_g$ orbitals \cite{Ran11,Ruegg11,Ruegg12,Ruegg13,Okamoto13,Okamoto14}
and pyrochlore iridate $j_{\rm eff}=1/2$ orbitals \cite{Fiete12,Kargarian13,QChen15} were proposed to exhibit
TIs as well as QAHIs with Chern number $C=1$. However, an important challenge which is faced in most single-TM based 
QAHIs proposed, is the fact that correlations need to be strong enough to drive magnetism and break time-reversal 
symmetry. At the same time, it should not be so strong so as to cause complete Mott localization.\cite{Okamoto14}
Recently, the [001] interface between TiO$_2$ and the half-semimetal CrO$_2$ has been proposed to form a QAHI, \cite{TCai15}
with higher Chern number $C=2$, but unfortunately exhibiting a small gap of $\sim 4$meV.

Taking a step forward, a promising class of materials is that of naturally occurring bi-component TM perovskite compounds with ordered
structure, i.e. the double perovskite (DP) compounds of general composition A$_{2}$BB$^{\prime}$O$_6$ with rock-salt ordering of 
two different TM atoms, B and B$^\prime$ \cite{SahaDasgupta14}. 
For example, in a $3d$-$5d$ or 3$d$-4$d$ DP, the 3$d$ TM  atoms (B) can
allow for a high energy scale for magnetism, while the 5$d$ or 4$d$ TM atoms (B$^\prime$) can feature strong SOC, making them ideal 
candidates in search for high temperature QAHIs. Physical separation of the ions which host magnetism from the ions which host 
strong SOC, as in the DPs, avoids the issue related to interplay of correlation effect and SOC at the same site, as in the case of 
single TM based candidates. Furthermore, in many of these 3$d$-5$d$ or 3$d$-4$d$ DP materials, the low-energy physics are typically described by $t_{2g}$-orbital 
bands of 5$d$ TM ions, which together with strong oxygen-covalency can strongly suppress the Jahn-Teller (JT) distortion. This
is in contrast to $e_g$-orbital based proposals which are prone to strong JT distortions that tend to favor trivial insulators.\cite{Pentcheva16,Haule}

In this article, we consider the case of Ba$_{2}$FeReO$_6$ (BFRO), which in bulk, is a half-metallic FM with high 
$T_c$ of $\approx$ 305K.\cite{Azimonte07} Here, the Ba, Fe and Re atoms 
occupy the A, B and B$^\prime$ sites respectively. Using first-principles density functional theory (DFT) calculations, 
combined with effective model calculations we show that an ultrathin [111] film of BFRO, consisting of layers of 
Fe/BaO$_3$/Re/BaO$_3$, forms a $C\!=\! 1$ QAH insulator with a large topological gap $\!\sim\! 100$meV and an estimated 
FM $T_c \sim 315$K. The large-gap and high $T_c$ should enable practical use of our proposal.

\begin{figure}[t]
\includegraphics[width=8.6cm,keepaspectratio]{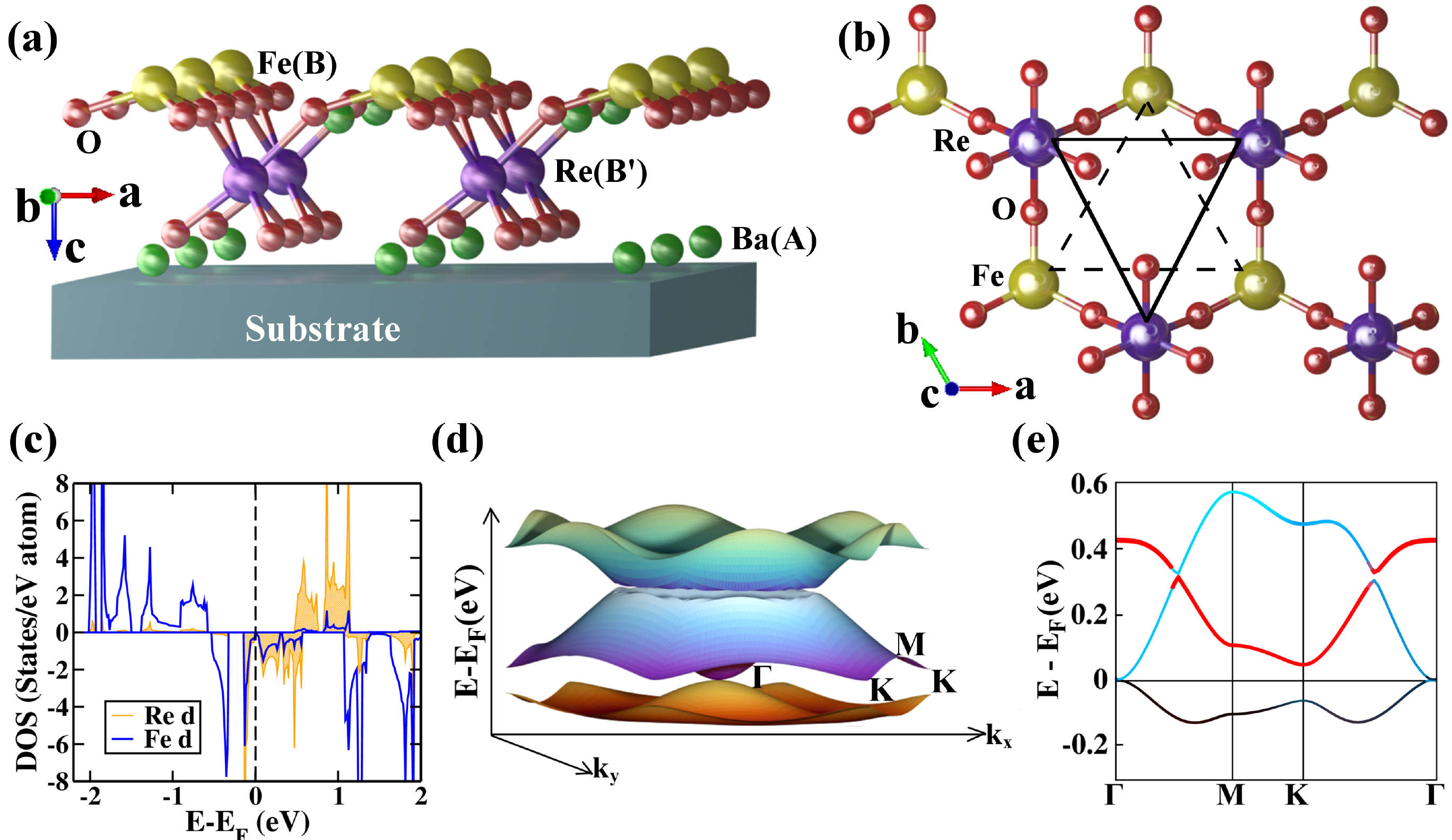}
\caption{(Color online) {\bf (a)}: Bilayer of Ba$_{2}$FeReO$_6$ (BFRO), with c-axis pointed along the [111] growth direction and terminated 
at Fe layer. {\bf (b)}: Buckled honeycomb network of Re and Fe atoms, viewed along the growth direction. Ba, Re, Fe and O atoms have been 
marked. In {\bf (b)} Ba atoms have been omitted for clarity. {\bf (c)}: Spin-polarized GGA density of states projected onto Fe 
and Re $d$ states. {\bf (d)}: The GGA band structure in minority spin channel plotted in the hexagonal BZ. {\bf (e)}: 
Same as in {\bf (b)}, but plotted along high symmetry directions; the colors 
(cyan/light gray, black, red/dark gray) show dominant orbital characters ($d_{xy}$, $d_{x^{2}-y^{2}}$, $d_{3z^{2}-r^{2}}$).}
\end{figure}

It is worth mentioning that our work goes well beyond earlier studies of DP films.\cite{Cook14,Baidya15} 
In particular our study identifies three key ingredients for formation of this QAHI, which should be broadly
applicable to other $t_{2g}$ physics dominated 3$d$-5$d$ or 3$d$-4$d$ half-metallic DPs like Sr$_{2}$FeMoO$_6$, 
Sr$_{2}$CrWO$_6$ with strong ferromagnetism\cite{Tokura98,Sarma03} driven by the 3$d$ ion:
\begin{description}
\item{(1)} Ultrathin films of such half-metallic oxides grown along the [111] direction show a strong trigonal distortion, favoring 
a non-Kramers doublet formed from 5$d$ or 4$d$ TM ion derived $t_{2g}$ orbitals 
in reduced $C_3$ crystal field symmetry. Following this, the band structure of the conducting 
spin channel features symmetry-protected linear Dirac band touchings at the 
hexagonal Brillouin zone (BZ) corners ($\pm K$)
and a single quadratic band touching (QBT) at the $\Gamma$-point.
\item{(2)} The breaking of in-plane inversion symmetries by localized 3$d$ TM ions gaps 
out the Dirac points at the BZ corners. The source of the gap opening at the Dirac
points is traced to a non-relativistic {\it orbital Rashba}-type effect.
For appropriate filling of the 5$d$/4$d$ ion, this leads to a half-semimetal, with the Fermi energy $E_F$ pinned to 
the QBT point. 
\item{(3)} The strong SOC of 5$d$/4$d$ TM ion gaps out the QBT.
For a filling where $E_F$ was pinned to the QBT point, this leads to a QAHI with a topological gap.
The strong SOC and the trigonal distortion also combine to pin the 5$d$/4$d$ TM moments to be perpendicular to the plane, 
leading to a high FM $T_c$.
\end{description}

\section{DFT calculations} 

Fig.1 shows the structure of an ultrathin [111] bilayer film of BFRO. The [111] growth direction 
results in the stacking of BaO$_{3}$, Re, BaO$_{3}$, Fe layers. The bilayer is terminated at the Fe layer facing vacuum, 
the bottom BaO$_3$ layer being supported by the substrate, as shown in Fig.1(a).\cite{Note1}
This results in three-fold oxygen coordination of Fe atom 
and six-fold oxygen coordination of Re atom, forming a buckled honeycomb lattice with two interpenetrating triangular lattices
of Fe and Re atoms, as shown in Fig.1(b). The atomic positions in the constructed thin film geometry were completely 
relaxed within the DFT optimization scheme, while the in-plane lattice constant of the film was held fixed at $a_0$,
with $a_{0} = 5.6922 \AA$ being the lattice constant for bulk BFRO, which is in almost perfect commensuration with the substrate
of the cubic BaTiO$_{3}$.\cite{Chen10} The buckled honeycomb lattice geometry for the [111] growth direction introduces a trigonal distortion in ReO$_6$ octahedra, signaled by $\approx 3.7^\circ$
deviation of O-Re-O bond angles from $90^\circ$. The DFT calculations employed for structural optimization as well as for
self-consistent calculation of electronic structure  were carried out within the generalized gradient approximation 
(GGA) \cite{Perdew96} for the exchange-correlation functional. We checked that our results were qualitatively unchanged upon
incorporating correlation effects within GGA+U.\cite{Liechtenstein95}
We used two choices of basis sets: (i) the plane-wave based 
pseudopotential method as implemented in the Vienna An-initio Simulation Package (VASP) \cite{VASP}
and (ii) the full potential linear augmented plane-wave method of Wien2k.\cite{Wien2k} GGA calculations were also 
carried out including SO for the Re 5$d$ states, which is a crucial ingredient to drive the QAHI.
For details of DFT calculations, see Appendix A.

\begin{figure}[t]
\includegraphics[width=8.6cm,keepaspectratio]{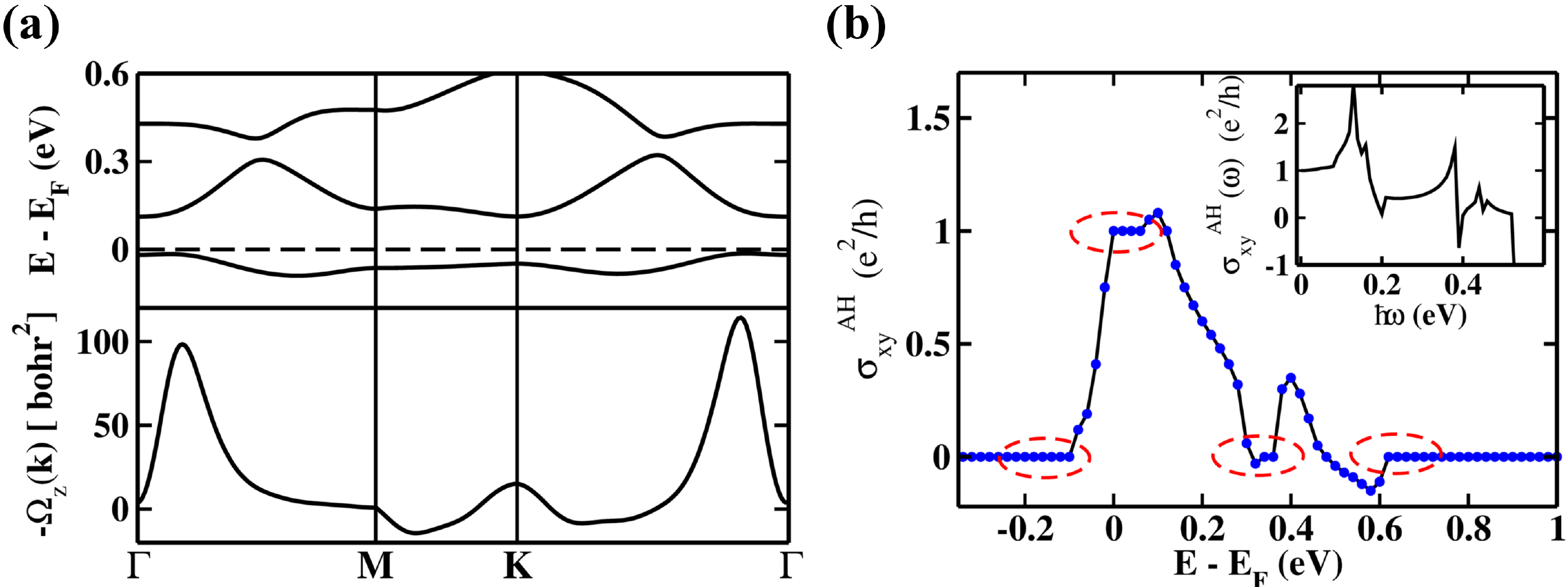}
\caption{(Color online) {\bf (a)}: GGA+SO band structure in minority spin channel (top panel) and calculated
Berry curvature (bottom panel) plotted along 
the high symmetry directions in the BZ. {\bf (b)}:  
Anomalous Hall conductivity of the band structure in {\bf (a)}, in
units of $e^2/h$. Quantized plateaus are highlighted. Inset: Frequency dependence of the real part of the optical 
AH conductivity $\sigma^{AH}_{xy}(\omega)$ in units of $e^2/h$.}
\end{figure}

Fig.1(c) shows the GGA spin-polarized DOS (without SO), projected onto Fe $d$ and Re $d$ states. The GGA results in a 
completely gapped solution in the majority spin channel, while a semimetallic state is obtained in the minority spin channel.
Our first important observation is that while the Fe $d$ states are completely occupied in the majority spin channel as 
in case of bulk DOS,\cite{Baidya15} in contrast to bulk DOS we also find
occupancy of Fe $d$ states in the minority spin channel, apart from the presence of strongly hybridized Fe-Re states close to
E$_F$. This suggests that Fe in the thin film geometry of BFRO to be in a nominal $d^{6}$ or 2+ state, as opposed to the 
nominal $d^{5}$ or $3+$ oxidation state in bulk BFRO, the latter being established by x-ray magnetic circular dichroism (XMCD) 
experiment \cite{Azimonte07} and by DFT calculations.\cite{Wu01,Baidya15}  Considering the nominal 2+ and 2- valences of Ba and O, this 
puts Re in a d$^{1}$ or 6+ valence state. This observation is validated by the computed magnetic moments of 2.99 $\mu_B$ 
(-0.39 $\mu_B$) at Fe (Re) site as compared to bulk values of 3.73 $\mu_B$ (-0.78 $\mu_B$).\cite{Baidya15} 
Nevertheless, the net moment in the cell 
is identical to the bulk value of $3.0 \mu_B$, with a significant fraction of the moment residing at oxygen sites in both cases 
due to the strong covalency effect. Thus, remarkably,
the reduction of dimensionality in moving from bulk to ultrathin
films, resulting in reduced O-coordination at Fe site, causes strong redistribution of charges between Fe 
and Re. As discussed below, this is one of the crucial aspect in defining the low energy electronic structure 
and the resulting QAHI.

We next focus on the details of the low energy minority spin band structure in the 2D Brillouin zone (BZ). This is found to
consist of three sheets as 
shown in Fig.1(d), dominantly consisting of the three $t_{2g}$ orbitals of Re, 
admixed with O $p$ and Fe $t_{2g}$ characters. The three-fold O-coordination on the Fe sites, leads to a highly reduced $C_3$
symmetry, as opposed to the bulk octahedral symmetry. 
This imprints itself as a strong trigonal distortion in the low energy band structure, splitting 
the Re $t_{2g}$ states into doubly degenerate $e_{g}^{\pi}$
and singly degenerate $a_{1g}$ state, with $a_{1g}$ state being energetically higher than $e_{g}^{\pi}$, as 
defined in the local octahedral 
coordinate system. In terms of global coordinates of the cell, they correspond
to linear combinations of $d_{xy}$/$d_{x^{2}-y^{2}}$, and $d_{3z^{2}-r^{2}}$. 

We find isolated gapped bands everywhere in the BZ, except at the $\Gamma$ point, where the valence and conduction bands touch 
quadratically [cf Fig. 1(d)]. Orbital projection of the band structure, as presented in Fig.1(e), shows that out of the three bands,
the $a_{1g}$ ($d_{3z^{2}-r^{2}}$) band remains completely empty. The other two bands exhibit a QBT at $\Gamma$,
one of them being fully occupied and another being empty, in order 
to satisfy the nominal $d^{1}$ filling of Re. Thus the [111] thin film geometry of 
BFRO with $d^{1}$ filling of Re, gives rise to a {\it half semi-metallic} band structure.

Starting with this GGA result, we next turn on the SOC. This induces an orbital moment of 0.13 $\mu_B$ at Re 
site, pointing opposite to the spin moment. As shown in Fig.~2(a),
the most dramatic effect of SOC on the band structure
is at the $\Gamma$ point, where it splits the QBT, opening up a gap of $\approx\! 100$ meV and
triggering a completely insulating state.
Using WANNIER90,\cite{Kunes10} which employs an
interpolation scheme, we have computed the Berry curvature $\Omega_z(\bk)$ over a fine $\bk$-mesh in the BZ.
Fig.2(b) plots $\Omega_z(\bk)$ along high symmetry directions,
showing that it is concentrated close to the $\Gamma$-point.
The anomalous Hall conductivity (AHC) obtained by integrating the Berry curvature over occupied bands in the
entire 2D BZ is plotted in Fig.2(b) as a function of varying energy. The nontriviality in band structure
shows up as quantized plateaus in this plot (circled for clarity).
We have confirmed this result using the Kubo-Greenwood formula  \cite{Kubo57} to obtain the
anti-symmetric optical conductivity $\sigma^{AH}_{xy}(\omega)$, shown in the inset of Fig.~2(b), 
with the AHC being given by the real part of $\sigma^{AH}_{xy}(\omega \to 0)$.
These calculations reveal that the insulating state of the [111] thin film of BFRO is a $C$ = 1 Chern 
insulator with a large gap $\approx 100$meV.

We have also computed the effective Fe-Fe magnetic exchange by total energy calculation of different 
magnetic configurations of Fe spins. For bulk BFRO, our results yield $J^{\rm bulk}_{\rm Fe-Fe} \approx 1$meV,
in excellent agreement with inelastic neutron scattering experiments \cite{Plumb13} (see Appendix B). 
For the [111] bilayer, we find a larger $J^{\rm film}_{\rm Fe-Fe} \approx 3.5$meV. Using this value,
we have carried out classical Monte Carlo simulations of an appropriate spin model of Heisenberg-Ising type. 
As discussed in Appendix B, we estimate the thin film FM ordering temperature $T_c$ to be $\approx$ 315K,
which is even slightly higher than the bulk $T_c$.

\begin{figure}[t]
\includegraphics[width=4cm]{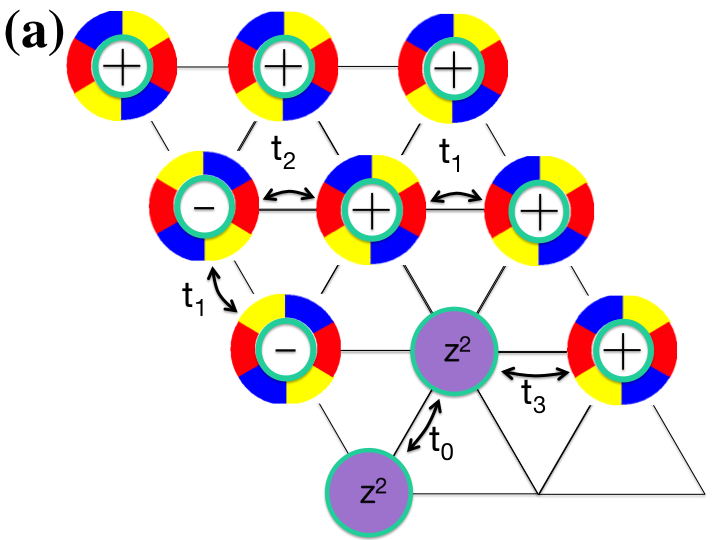}
\includegraphics[width=4cm]{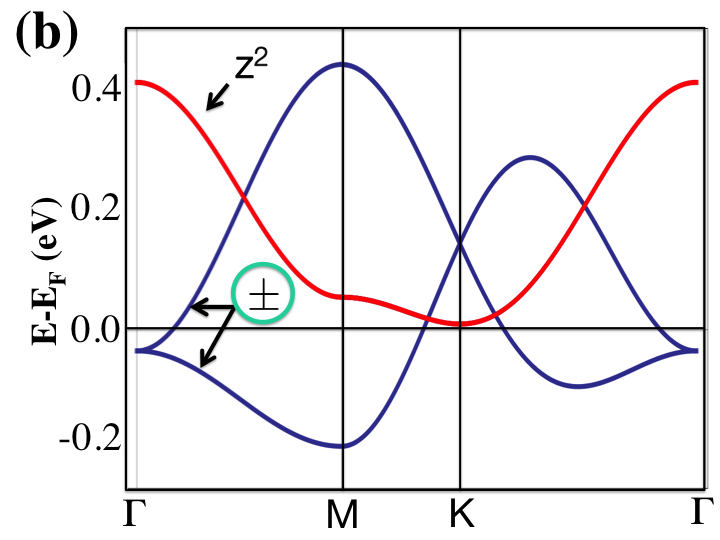}
\caption{(Color online) {\bf (a)} The intraorbital ($t_0, t_1$) and interorbital ($t_2, t_3$) hoppings on
the triangular Re lattice between $t_{2g}$ orbitals with angular
momenta $m_\ell\!=\! 0$ (denoted $z^2$) and $m_\ell\!=\! \pm 1$. For $m_\ell \!=\! \pm 1$,
the wheels depict opposite phase windings. {\bf (b)} 
Decoupled $m_\ell\!=\! 0$ and coupled $m_\ell\!=\!\pm 1$ bands for $(\Delta_{\rm trg},t_0,t_1,t_2) \!=\! (60, 45, 20, 80)$~meV, 
where all other terms in Hamiltonian have been set to zero (see text).}
\label{fig:lattice}
\end{figure}

\section{Effective model} 

In order to gain microscopic understanding, we carry out further analysis in terms of a tight-binding effective model for the
three minority-spin bands close to E$_F$. We work with 
eigenstates of $L^{\rm eff}_z$, labeled as
$|m^{\rm eff}_\ell \ra$, which diagonalize the trigonal crystal field Hamiltonian, $\hat{z}$ being the [111] direction.
In global coordinates,
$|0 \ra \! \equiv\! |d_{3 z^2 - r^2} \ra$ is at higher energy, while
$|\pm 1 \ra  \!\equiv\! \frac{1}{\sqrt{2}}(|d_{x^2-y^2}\ra \pm i |d_{xy}\ra)$ form a low energy non-Kramers doublet (see Appendix C).
The $|m^{\rm eff}_\ell\ra$ orbitals are schematically shown in Fig.3(a), with $z^2 \!\equiv\! |0 \ra$ depicted as filled circles and 
the $|\pm 1\ra$ orbitals having opposite phase windings of $\mp 2\pi$ corresponding to the appropriate real space wavefunctions, 
shown by color wheels in the figure.
We set $\Psi^\dagger_\bk \!=\! (c_{\bk\uparrow0}^\dagger, c_{\bk\uparrow+}^\dagger,c_{\bk\uparrow-}^\dagger)$ in the $m_\ell$-basis, 
and consider $H_{\rm eff} \!=\! \sum_\bk \Psi^\dagger_\bk {H}^\pdg_\bk \Psi^\pdg_\bk$. The intraorbital [$\varepsilon(\bk)$] and interorbital 
[$\eta(\bk)$] dispersions in ${H}^\pdg_\bk$ are given by,
\begin{eqnarray*}
\!\!\!\!\! \varepsilon^\pdg_0(\bk) \!\!&=&\!\! -2 t_0\! (\cos k_1 \!+\! \cos k_2 \!+\! \cos k_3) \\
\!\!\!\!\! \varepsilon^\pdg_\pm (\bk) \!\!&=&\!\! -2 t_1\! (\cos k_1 \!+\! \cos k_2 \!+\! \cos k_3) \!-\!\! \Delta_{\rm trg} \! \pm \! \gamma^{R}_\bk \! \mp \! \frac{\lambda}{2}\\
\!\!\!\!\! \eta^\pdg_{+-}(\bk) \!\!\!&=&\!\!\! 2 t_2 (\cos k_1 \!+\! \omega \cos k_2 \!+\! \omega^2 \cos k_3)\\
\!\!\!\!\! \eta^\pdg_{0+}(\bk) \!\!\!&=&\!\!\! \eta^*_{0-}(\bk)\!=\! 2 t_3 (\cos k_1 \!+\! \omega \cos k_2 \!+\! \omega^2 \cos k_3),
\end{eqnarray*}
where  $k_1=k_x$, $k_2=-k_x/2+k_y \sqrt{3}/2$, $k_3=-k_1-k_2$, and  $\omega={\rm exp}(i 2 \pi/3)$. Here we have set
$t_0,t_1$ are intraorbital hopping amplitudes, while the 
interorbital $t_{2,3}$ are complex hoppings due to the phase winding of the $|\pm 1 \ra$ wavefunctions, as shown in Fig. 3(a).
$\Delta_{\rm trg}$ represents
the trigonal splitting. Due to the spin polarized character
of these bands, SOC acts as a Zeeman field on the orbital pseudospins via $- \lambda L^{\rm eff}_z S_z$, 
$\lambda$ being the strength of the SOC. Finally,
$\gamma^{R}_\bk \!=\! 2 t_R (\sin k_1 \!+\! \sin k_2 \!+\! \sin k_3)$ encodes an 
{\it orbital Rashba}-type effect due to the presence of Fe atoms which break inversion symmetry
of the Re lattice, as discussed below.

\begin{figure}[t]
\includegraphics[width=7cm]{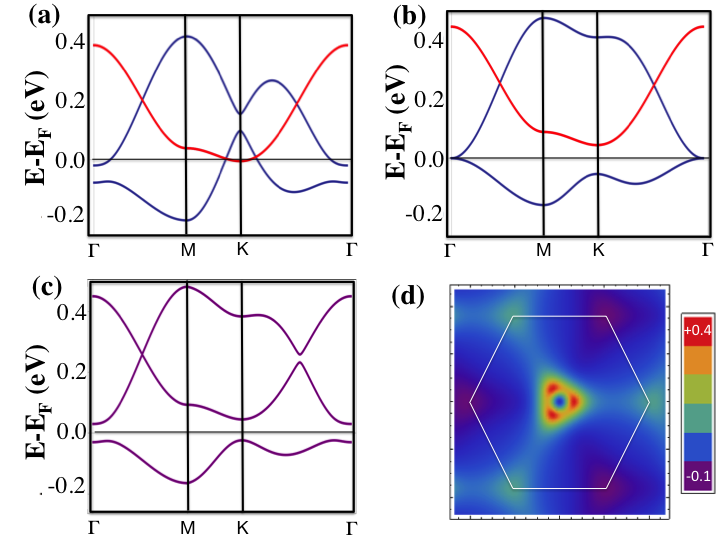}
\caption{(Color online) Band structure obtained by setting {\bf (a)} $(t_R \!=\! 0, \lambda \! \ne\! 0)$,  {\bf (b)} $(t_R \! \ne\! 0, \lambda\! =\! 0)$, and
{\bf (c)} $(t_R \!\ne\! 0, \lambda \!\ne\! 0)$ in the effective Hamiltonian (see text). {\bf (d)} Berry curvature 
$\Omega_z(\bk)/2\pi$ for the lowest band, with parameters as in (c), in units where the Re triangular lattice constant is set to unity.
The Berry curvature is peaked slightly away from the $\Gamma$ point, and is asymmetric under
$\bk \to -\bk$ due to broken inversion.}
\label{fig:disp}
\end{figure}

\begin{figure}[b]
\includegraphics[width=8cm]{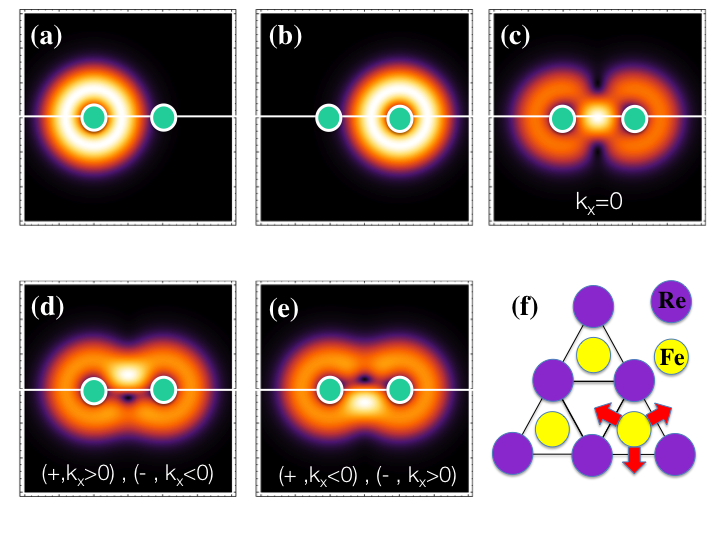}
\caption{(Color online) {\bf (a,b)} Electron density for localized pseudospin states $m_\ell=\pm 1$ shown at neighboring Re sites 
$\br$ and $\br+\hat{x}$.
{\bf (c-e)} Bond electronic density obtained by superposing electron density of the same orbital type (both $m_\ell=+1$ or 
both $m_\ell = -1$) for $k_x=0$, and $k_x \neq 0$, illustrating electric dipole ${\cal P}_y \sim m^{\rm eff}_\ell \sin k_x$, which changes
sign under reversing sign of $m^{\rm eff}_\ell$ or $k_x$.
{\bf (f)} Triangular lattice structure of Re atoms and underlying Fe atoms.
Arrows schematically indicate the in-plane electric fields, generated by symmetry-breaking due to Fe atoms, which
can linearly couple to $\vec {\cal P}$.}
\label{fig:dipole}
\end{figure}

To clarify the effect of various terms in the effective Hamiltonian, we plot the band structure setting
$t_R\!=\!\lambda\!=\! t_3 \!=\! 0$ and fixing $(\Delta_{\rm trg},t_0,t_1,t_2) \!=\! (60, 45, 20, 80)$~meV. As seen from Fig.~3(b), this
broadly reproduces the GGA bandwidth. In this limit, the $|0\ra$ band is completely decoupled, while the coupled $|\pm 1 \ra$ bands 
exhibit a $C_3$ and ${\cal T}$ symmetry protected QBT at $\Gamma$ 
as well as inversion ${\cal I}$ and ${\cal T}$ symmetry protected linear Dirac band crossings at 
$\pm K$, ${\cal T}$ being the time-reversal for 
spinless fermions (as appropriate for a spin-polarized half-metal).

Turning on SOC, $\lambda \neq 0$, breaks ${\cal T}$ and gaps out these band touchings, as illustrated in Fig.4(a) for $\lambda=30$meV.
The resulting bands derived from $|\pm 1\ra$ are topologically nontrivial, with Chern numbers $C=\pm 1$; however, the strong dispersion leads to a 
metal and not a QAHI. 

On the other hand, setting $t_R \neq 0$ only breaks ${\cal I}$. As shown in Fig.~4(b) for $t_R=45$meV, this only
gaps out the Dirac points at $\pm K$, but
preserves the QBT at $\Gamma$ point, perfectly reproducing the GGA dispersion [cf. Fig.1(e)].
To understand the physical origin of $t_R$, we plot in Fig.~5(a,b) the ring-like
probability density for the $|\pm 1\ra$ states localized on a 
pair of neighboring Re sites, positioned at $\br$ and  $\br+\hat{x}$, where $\hat{x}$ is a Re triangular lattice bond direction.
Remarkably, superposing these two localized states
with a relative phase factor of ${\rm e}^{i k_x}$, to form a Bloch state at momentum $k_x$, leads to a
probability density distribution which features a ${\bf k}$-dependent bond electric dipole moment
${\cal P}_y \sim m^{\rm eff}_\ell \sin k_x$. This `Bloch dipole' ${\cal P}_y$ vanishes for $k_x=0$ (see Fig.~5(c)), and
changes sign under reversing the direction of $\bk$ or changing $m^{\rm eff}_\ell$ from $+1$ to $-1$ (or vice versa), as shown 
in Figs.~5(d,e). On the triangular Re lattice, this leads to a $\bk$-dependent `Bloch dipole' 
which can linearly couple to the electric fields arising due to the presence of Fe atoms breaking the inversion 
symmetries of the triangular Re lattice, as schematically shown by arrows 
in Fig.5(f). This gives rise to the term $\gamma^{R}_\bk$ in the effective Hamiltonian, as shown in Appendix D.  A closely related effect has been 
shown to account for the spin-orbital texture of surface bands in other materials,\cite{CKim11,JHHan14}
though unlike those cases, here the effect arises from in-plane inversion breaking.

Finally, with $t_R \neq 0$ and $\lambda \neq 0$, we are led 
to a QAHI, with the lowest band having $C$ = 1 [cf Fig.4(c)], in agreement with the DFT results. The Berry
curvature of the lowest band $\Omega_z(\bk)$, shown in Fig.~4(d), is concentrated around the $\Gamma$-point
and is in 
semi-quantitative agreement with the DFT results. The inversion symmetry breaking is reflected in
$\Omega_z(\bk) \neq \Omega_z(-\bk)$.

\section{Conclusions}

Using a combination of first-principles calculations and analysis of an effective Hamiltonian, we 
have addressed the challenge of possible material realization with a large band-gap, high-temperature 
Quantum Anomalous Hall insulating state. We propose that ultra-thin films of mixed row (3$d$-5$d$ or 3$d$-4$d$) 
double perovskites with [111] orientation have (a) 3$d$ TM ions providing strong magnetism, 
(b) 5$d$ or 4$d$ TM ions with partially filled $t_{2g}$ bands with a SOC,
which make them strong contender for the realization of QAHI. 
We note that several of these DPs are rare examples of materials with half-metallic ground
state with above room temperature Curie temperature, justifying this choice.\cite{Tokura98,Sarma03,Azimonte07} 
We have demonstrated this for Ba$_{2}$FeReO$_6$ through rigorous first-principles
DFT calculations, corroborated with analysis of a model Hamiltonian. 
We show that ultrathin [111] film of BFRO is indeed a robust
high temperature ($T_c \approx$ 315K), large gap ($\approx$ 110 meV), $C=1$ QAHI, opening 
up the possibility of room temperature applications in dissipationless electronic circuit. 
Edge interfaces of such oxide QAHI states with correlated superconductors might lead to high 
temperature realization of chiral Majorana edge states.
\cite{KLWang2016}.
We have uncovered the key role of the orbital Rashba-type effect, which works hand-in-hand with SOC
to drive the nontrivial QAHI. This orbital Rashba-type effect is expected 
to play an important role in a broader class of phenomena in orbitally degenerate oxides. 

\section{Acknowledgement}

S.B. and T.S.D thank Department of Science and Technology, India for the computational facility through Thematic Unit  
of Excellence of Nano-mission. AP thanks A. M. Cook for preliminary
discussions and related collaboration, and acknowledges funding from NSERC of Canada, and support from
the Aspen Center for Physics through National Science Foundation grant PHY-1066293. 
UVW acknowledges support from a 
Thematic Unit for Computational Materials Science at JNCASR, and a JC Bose National Fellowship of the department
of science and technology, government of India.

\appendix
\section{Details of DFT Methodology}

Our first-principles based density functional theory (DFT) calculations were carried out in the generalized gradient approximation (PBE-GGA)
\cite{Perdew96, PAW1, PAW2} of the exchange-correlation potential with a $13\times13\times2$ k-point grid. The computations were performed in two different ways : (i) using the all-electron full-potential linear augmented plane-wave (FP-LAPW) method as implemented in the WIEN2k code,
\cite{Wien2k} and (ii) using a plane-wave-based pseudopotential method as implemented in the Vienna \textit{ab initio} Simulation Package 
(VASP).\cite{VASP} Close results were obtained from two different approaches. The spin-orbit coupling was included in the fully-relativistic schemes. 

The muffin-tin radii (R$_{MT}$), used in FP-LAPW calculations, were $2.50$, $1.80$, $1.86$ and $1.60$ a.u. for Ba, Fe, Re and O, respectively. To achieve reasonable accuracy for ground state in the FP-LAPW calculation, 
the number of plane waves was set by the criteria, K$_{max}$.R$_{MT}=7$ where
K$_{max}$ is the plane wave cut off value and R$_{MT}$ is the smallest muffin-tin radius. 
The energy cutoff 600 eV was chosen for calculations in plane wave basis set. The self-consistent convergence was tested with choices of different k-point grids. The ionic position relaxation was carried out in both FP-LAPW and plane wave calculations. For structural optimization 
a force convergence of 0.01eV/\AA~ was ensured in plane wave basis calculations. On the other hand in FP-LAPW calculations, optimization of the atomic positions 
were performed using force convergence of 0.001 Ry.au$^{-1}$.

The non-trivial properties were calculated by constructing maximally-localized Wannier functions (MLWFs) for the selected three bands close to Fermi level possessing Re $t_{2g}$ Bloch orbitals character (confirmed from orbital character plotting) using WANNIER90 code \cite{Wannier90} interfaced with WIEN2k code. Starting from DFT generated ground state eigenvalue, eigenstate and Hamiltonian matrices, WANNIER90 is capable of constructing MLWFs corresponding to selected energy bands.\cite{Kunes10} The Berry curvature matrix was calculated for whole Brillouin Zone in the basis of MLWFs. The details of wannierization scheme is available in reference \cite{Vanderbilt06}. The intrinsic Anomalous Hall Conductivity was calculated by integrating Berry curvature over dense k mesh of $370\times370\times65$ on whole Brillouin zone (BZ) using WANNIERE90 code. Convergence of AHC value was checked for different choices of dense k-mesh over whole BZ. Further the calculations of ac anomalous Hall conductivity was carried out using Kubo-Greenwood formula for optical conductivity using WANNIER90 code. It gives ac optical conductivity which has both symmetric and anti-symmetric part. 
The anti-symmetric part gives ac anomalous Hall conductivity which for zero optical frequency gives intrinsic anomalous Hall conductivity (AHC). The details of calculating ac optical conductivity can be obtained in WANNIER90 code \cite{Wannier90}.

\section{Finite temperature ferromagnetic transition of thin film: Monte Carlo study}

\subsection{Bulk exchange}

The DFT calculation of total energy using various Fe magnetic configurations shows that the effective ferromagnetic exchange between Fe moments in bulk BFRO is  $J_{FF} = 0.97$ meV. Considering an effective model of only Fe spins (integrating out
the Re electrons), this implies that flipping a single spin in the ferromagnetic state leads to an energy cost 
$E_{\rm flip}=2 J_{FF} S_F^2 z_{\rm fcc}$ where $z_{\rm fcc}=12$ is the coordination number on the fcc lattice of Fe moments
and $S_F$ is the effective Fe spin.
Alternatively the magnetic exchange can be computed from a spin model where both Fe and Re are retained in the model.
Within such a picture, the effective description of the magnetism is 
a Heisenberg model on the Re-Fe lattice, with an antiferromagnetic exchange coupling $J_{FR}$, which describes ferrimagnetic
order with unequal Fe and Re moments. Within such a model, the energy cost to flipping a spin is 
$E_{\rm flip}=2 J_{FR} S_F S_R  z_{\rm c}$ where the cubic lattice coordination number $z_c=6$ and $S_R$ denotes
the effective Re spin.  Equating the two expressions, we find $J_{FF} = J_{FR} (S_R/S_F) (z_c/z_{\rm fcc})$. Using
$J_{FR} \approx 3$meV and $S_F/S_R \approx 1.6$
from inelastic neutron scattering experiments \cite{Plumb13} on bulk BFRO, we find $J_{FF} \approx 0.94$ meV,
remarkably close to the DFT estimate.

\subsection{Thin film model}
\begin{figure}[tbh]
\includegraphics[width=8.6cm,keepaspectratio]{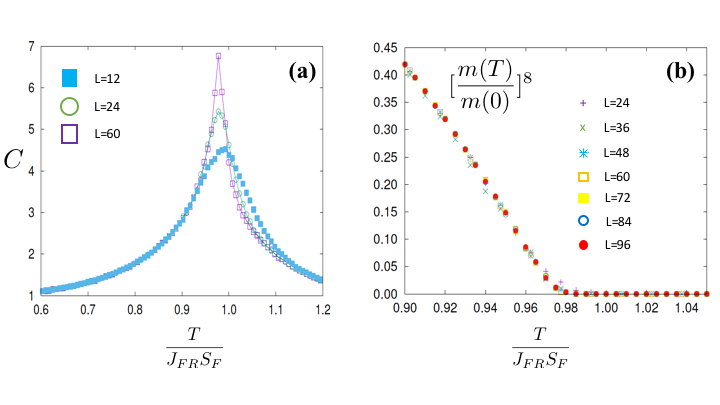}
\caption{(Color online) Temperature dependence of {\bf (a)} specific heat per unit cell of the honeycomb lattice (in units of $k_B$),
and {\bf (b)} magnetization per site of the honeycomb lattice (normalized to its value at $T=0$), for the
Heisenberg-Ising model in Eq.~1, for illustrative system sizes containing
$2 L^2$ spins. We have plotted $[\frac{m(T)}{m(0)}]^{1/\beta}$ to illustrate data collapse for an Ising order
parameter exponent $\beta=1/8$.
Using these simulations we determine the FM transition temperature $T_c/S_F J_{FR} \approx 0.975(5)$.}
\label{fig:MC}
\end{figure}

 In the thin film geometry, our DFT computation yields $J_{FF} \! \approx\! 3.5$meV, larger than the bulk
value. Proceeding as in bulk case,
this means flipping a single Fe spin in the ferromagnetic state costs an 
energy $E_{\rm flip}\!=\! 2 J_{FF} S_F^2 z_{\triangle}$, where $S_F\!=\! 2$ and the triangular lattice coordination number 
$z_{\triangle}\!=\! 6$. 
Considering the alternative Fe-Re model, we note here, unlike in bulk, the strong trigonal splitting and large SOC conspire to pin 
the Re moments to lie perpendicular to the film as discussed in the paper, so they effectively 
behave as Ising moments $\sigma=\pm 1$.
The effective model takes the form
\be
H_{\rm Fe-Re} = J_{FR} \sum_{i \in {\rm Fe}, \delta} S^z_i \sigma_{i+\delta}
\ee 
Note that the Fe spins are still vector spins, only the exchange is highly anisotropic. In the ferrimagnetically
ordered state, flipping an Fe spin now costs energy $E_{\rm flip} = 2 J_{FR} S_F |\sigma|  z_H$, where $|\sigma|=1$ and
$z_H=3$ is the honeycomb lattice coordination number. Equating the two expressions,
$J_{FR} = S_F J_{FF} z_{\triangle}/z_H \approx 14$meV.
We have carried out Monte Carlo simulations of this mixed Heisenberg-Ising model on the honeycomb lattice.
Fig.~\ref{fig:MC} shows the specific heat divergence [panel (a)] of this model at the paramagnetic to ferromagnetic transition
and the order parameter [panel (b)] for a system size of $2 L^2$ spins. We have plotted the scaled magnetization
$(\frac{m(T)}{m(0)})^8$ corresponding to an Ising exponent $\beta=1/8$ to illustrate its linear behavior near $T_c$
and data collapse for various system sizes.
 From such simulations, we estimate
$T_c/S_F J_{FR} \approx 0.975(5)$, which yields $T_c \approx 315$K, which we quote in the paper.
Use of a quantum renormalized  spin value $S_F^2 \to S_F (S_F+1)$ would suggest an even higher estimate for $T_c$.
We have also carried out a simulation of an effective triangular lattice model of vector Fe moments coupled via
Ising-exchange interactions (mediated by pinned Re moments), finding very similar results. Our result for the 
high FM $T_c$ of the thin film is thus robust.

\section{Orbital wavefunctions}

In the frame of the local octahedral coordinates $(x',y',z')$, we can 
express the $t_{2g}$ wavefunctions in terms of angular momentum
eigenstates split by the local trigonal distortion as
\bea
|a_{1g} \ra &=& \frac{1}{\sqrt{3}} (|y'z'\ra + |z'x'\ra + |x'y'\ra) \\
|e^+_{1g} \ra &=& \frac{1}{\sqrt{3}} (\omega^2 |y'z'\ra + \omega |z'x'\ra + |x'y'\ra) \\
|e^-_{1g} \ra &=& \frac{1}{\sqrt{3}} (\omega |y'z'\ra + \omega^2 |z'x'\ra + |x'y'\ra)
\eea
where, $\omega={\rm exp}(i 2 \pi/3)$. Here, $|a_{1g}\ra$ refers to the singlet state with $\vec L_{\rm eff} \cdot \hat{n}_{111} = 0$, while
$e^{\pm}_{1g}$ have $\vec L_{\rm eff} \cdot \hat{n}_{111} = \pm 1$. In the text, we work with
global coordinates $(x,y,z)$ such that on the Re sites,
\bea
x' &=& - \sqrt{\frac{2}{3}} y + \frac{1}{\sqrt{3}} z \\
y' &=& \frac{1}{\sqrt{2}} x  + \frac{1}{\sqrt{6}} y  + \frac{1}{\sqrt{3}} z \\
z' &=& - \frac{1}{\sqrt{2}} x  + \frac{1}{\sqrt{6}} y  + \frac{1}{\sqrt{3}} z
\eea
In these global coordinates, the 
\bea
|a_{1g} \ra &= & 3 z^2-r^2 \\
|e^\pm_{1g} \ra &\propto & |x^2-y^2\ra \pm  i |x y\ra - \frac{1}{\sqrt{2}} (|yz\ra \pm i |zx\ra)
\eea
Projecting to the $z=0$ plane of the triangular lattice, 
the last term does not play an important role, and we thus arrive at the $e_{1g}$ doublet being dominantly
given by $d_{x^2-y^2} \pm i d_{xy}$, having angular momentum projection
perpendicular to the bilayer given by $L^z_{\rm eff}=\pm 1$.

\section{{\it Orbital Rashba}-type effect}

\subsection{1D dispersion}
 As discussed in the main text of the paper, formation of a Bloch state with
momentum $\bk$ out of the $m^{\rm eff}_\ell=\pm 1$ states results into a nonzero electric dipole moment.
For one-dimensional motion, for example along
$\hat{x}$, this leads to ${\cal P}_y = \alpha m^{\rm eff}_\ell \sin k$, where $\alpha$ is a proportionality constant and $k \equiv k_x$.
If inversion symmetry is broken in the system,
such that there is an electric field ${\cal E}_y \neq 0$, this leads to an energy correction $- {\cal P}_y {\cal E}_y$
in the Hamiltonian which shifts the energy of the Bloch state, with the energy shift 
changing sign for $m^{\rm eff}_\ell \to -m^{\rm eff}_\ell$ or reversing the direction of $\bk$, 
so that time-reversal is preserved. This can be captured with the Hamiltonian
\bea
H_R = - i \frac{t_R}{2} \sum_{r, m=\pm 1} m (c^\dg_{r, m} c^\pdg_{r+x, m} - c^\dg_{r+x,m} c^\pdg_{r,m})
\eea
where $r$ is the site index.
In momentum space representation this becomes, 
\be
H_R = 2 t_R \sum_{k,m=\pm 1} {\rm sgn}(m) \sin k ~ c^\dg_{k,m} c^\pdg_{k,m},
\ee
where $t_R = \alpha {\cal E}_y$. 

\subsection{2D triangular lattice}
Moving from 1D to the 2D triangular lattice, which is the relevant lattice in this study, each bond on the triangular lattice has a nonzero
electric dipole generated by the Bloch state with momentum $\bk$. Now, the set of Fe atoms generates a nonzero electric
field perpendicular to the bonds of the Re triangular lattice as depicted in Fig.~5(f) of the paper. In this case, for each
bond direction, one gets a coupling of the type discussed in the context of 1D dispersion, leading to the Hamiltonian
\bea
H = - i \frac{t_R}{2} \sum_{\br, m=\pm 2,\delta} m (c^\dg_{\br, m} c^\pdg_{\br+\delta, m} - c^\dg_{\br+\delta,m} c^\pdg_{\br,m})
\eea
where $\delta=\hat{x}, -\hat{x}/2+\sqrt{3}\hat{y}/2, -\hat{x}/2-\sqrt{3} \hat{y}/2$ denote the three directions on the triangular lattice. 
In momentum space representation, this yields the term discussed in the paper,
\bea
H = \sum_{\bk, m=\pm 2,\delta} {\rm sgn}(m) \gamma^{R}_\bk c^\dg_{\bk, m} c^\pdg_{\bk, m}
\eea
where $\gamma^{R}_\bk = 2 t_R (\sin k_1 + \sin k_2 + \sin k_3)$, with $k_1=k_x$, $k_2=-k_x/2+k_y \sqrt{3}/2$, $k_3=-k_1-k_2$.
Since this is like a Rashba interaction, but with orbital degrees of freedom instead of spin,
and with the electric field pointing in the plane rather than out of the plane, we refer to it as an {\it orbital Rashba}-type effect.
The full Hamiltonian must also allow for electric fields pointing out to the plane, which would allow for
additional orbital Rashba terms \cite{CKim11,JHHan14} which would drive a conventional Rashba effect in the presence of SOC. 
A comparison of model calculations with our DFT results 
however show that the effect of such Rashba terms are small.

\vskip 2in
$\ast$ Presently at  Center for Correlated Electron Systems, Institute for Basic Science (IBS),
Seoul 151-742, Republic of Korea; Department of Physics and Astronomy, Seoul National University (SNU), Seoul, 151-742, Republic of Korea.\\
$\dagger$ Email: t.sahadasgupta@gmail.com


\end{document}